\documentclass{aa}

\usepackage{graphicx}
\usepackage{txfonts}
\usepackage[flushleft]{threeparttable} %For table footnotes

\usepackage{hyperref}
\hypersetup{colorlinks=true,linkcolor=blue,citecolor=blue,filecolor=blue,urlcolor=blue}

\usepackage{multirow}

\newcommand{\hei}{He~{\sc i}}
\newcommand{\heii}{He~{\sc ii}}

\def\lesssim{\mathrel{\hbox{\rlap{\hbox{\lower4pt\hbox{$\sim$}}}\hbox{$<$}}}}

\def\gtrsim{\mathrel{\hbox{\rlap{\hbox{\lower4pt\hbox{$\sim$}}}\hbox{$>$}}}}

\begin{document}

   \title{Optical and near-infrared spectroscopy of the black hole transient 4U 1543--47 during its 2021 ultra-luminous state}

   \author{J. Sánchez-Sierras\inst{\ref{i1},\ref{i2}}
          \and
          T. Muñoz-Darias\inst{\ref{i1},\ref{i2}}
          \and
          J. Casares\inst{\ref{i1},\ref{i2}}
          \and
          G. Panizo-Espinar\inst{\ref{i1},\ref{i2}}
          \and
          M. Armas Padilla\inst{\ref{i1},\ref{i2}}
          \and
          J.~Corral-Santana\inst{\ref{i3}}
          \and
          V. A. Cúneo\inst{\ref{i1},\ref{i2}}
          \and
          D. Mata Sánchez\inst{\ref{i1},\ref{i2}}
          \and
          S. E. Motta\inst{\ref{i4},\ref{i7}}
          \and
          G. Ponti\inst{\ref{i4},\ref{i5}}
          \and
          D. Steeghs\inst{\ref{i9}}
          \and
          M.A.P. Torres\inst{\ref{i1},\ref{i2}}
          \and
          F. Vincentelli\inst{\ref{i1},\ref{i2}}
          }

   \institute{Instituto de Astrofísica de Canarias, E-38205 La Laguna, Tenerife, Spain \label{i1}
         \and
             Departamento de Astrofísica, Universidad de La Laguna, E-38206 La Laguna, Tenerife, Spain \label{i2}
         \and
             European Southern Observatory, Alonso de Cordova 3107, Vitacura, Casilla 19001, Santiago de Chile, Chile \label{i3}
         \and
             Osservatorio Astronomico di Brera, Via E. Bianchi 46, I-23807 Merate (LC), Italy \label{i4}
         \and
             Department of Physics, University of Oxford, Denys Wilkinson Building, Keble Road, Oxford OX1 3RH, UK \label{i7}
         \and
             Max-Planck-Institut fur Extraterrestrische Physik, Giessenbachstrasse, D-85748, Garching, Germany \label{i5}
         \and
             Department of Physics, University of Warwick, Coventry CV4 7AL, UK \label{i9}
             }

   \date{Received 13th Dec 2022; Accepted 13th Mar 2022}
\titlerunning{OIR spectroscopy during the ultra-luminous state of 4U 1543--47}
\authorrunning{Sánchez-Sierras et al. 2022}
% \abstract{}{}{}{}{} 
% 5 {} token are mandatory
 
  \abstract
  {
    We present simultaneous optical and near-infrared spectra obtained during the 2021 outburst of the black hole transient 4U 1543–47. The X-ray hardness-intensity diagram and the comparison with similar systems reveal a luminous outburst, probably reaching the Eddington luminosity, as well as a long-lasting excursion to the so-called ultra-luminous state. VLT/X-shooter spectra were taken in two epochs 14 days apart during the early and brightest part of the outburst, while the source was in this ultra-luminous accretion state. The data show strong H and \ion{He}{i} emission lines, as well as high-excitation \ion{He}{ii} and \ion{O}{iii} transitions. Most lines are single-peaked in both spectra, except for the \ion{O}{iii} lines that exhibit evident double-peaked profiles during the second epoch. The Balmer lines are embedded in broad absorption wings that we believe are mainly produced by the contribution of the A2V donor to the optical flux, which we estimate to be in the range of 11 to 14 per cent in the $r$ band during our observations. Although no conspicuous outflow features are found, we observe some wind-related line profiles, particularly in the near-infrared. Such lines include broad emission line wings and skewed red profiles, suggesting the presence of a cold (i.e. low ionisation) outflow with similar observational properties to those found in other low-inclination black hole transients.
    }


   \maketitle

%--------------------------------------------------------------------
%--------------------------------------------------------------------

%-------------------------------------------------------------------
%- INTRODUCTION ----------------------------------------------------
%-------------------------------------------------------------------
\section{Introduction}

X-ray binaries (XRBs) are stellar systems where a compact object, a black hole (BH) or a neutron star, is accreting mass from its donor star. BHs are found almost exclusively as transient objects, the so-called BH transients, which remain in a low-luminosity state most of their lives (e.g. \citealt{Done2007,Belloni2011,Corral-Santana2016}). This quiescent period can be sporadically interrupted by outburst events characterised by a dramatic increase in luminosity across the electromagnetic spectrum, when a complex combination of accretion and outflow processes takes place (e.g. \citealt{Fender2016}). Two largely different accretion modes are observed during the outburst. The hard state is associated with the early and late phases of the outburst, and characterised by a hard, power-law-like X-ray spectrum and the presence of a steady compact radio jet. By contrast, a thermal disc component dominates the X-ray spectrum during the soft state, when the radio jet is quenched \citep[see e.g.][]{Fender2004, Russell2011}. In addition to intermediate states in between the previous two, a particularly luminous state characterised by a soft energy spectrum and a steep power law has been identified in some sources (e.g. GRO J1655--40; \citealt{Motta2012}), the so-called ultra-luminous state (also known as the anomalous state; see e.g. \citealt{Belloni2016} and references therein). The above accretion states can also be identified in neutron star systems (e.g. \citealt{Munoz-Darias2014}) and to some extent in active galactic nuclei (e.g. \citealt{FernandezOntiveros2021}). Likewise, they have also been proposed to be present in tidal disruption events (e.g. \citealt{Wevers2021}).

\begin{figure*}
    \centering
    \includegraphics[width=17truecm]{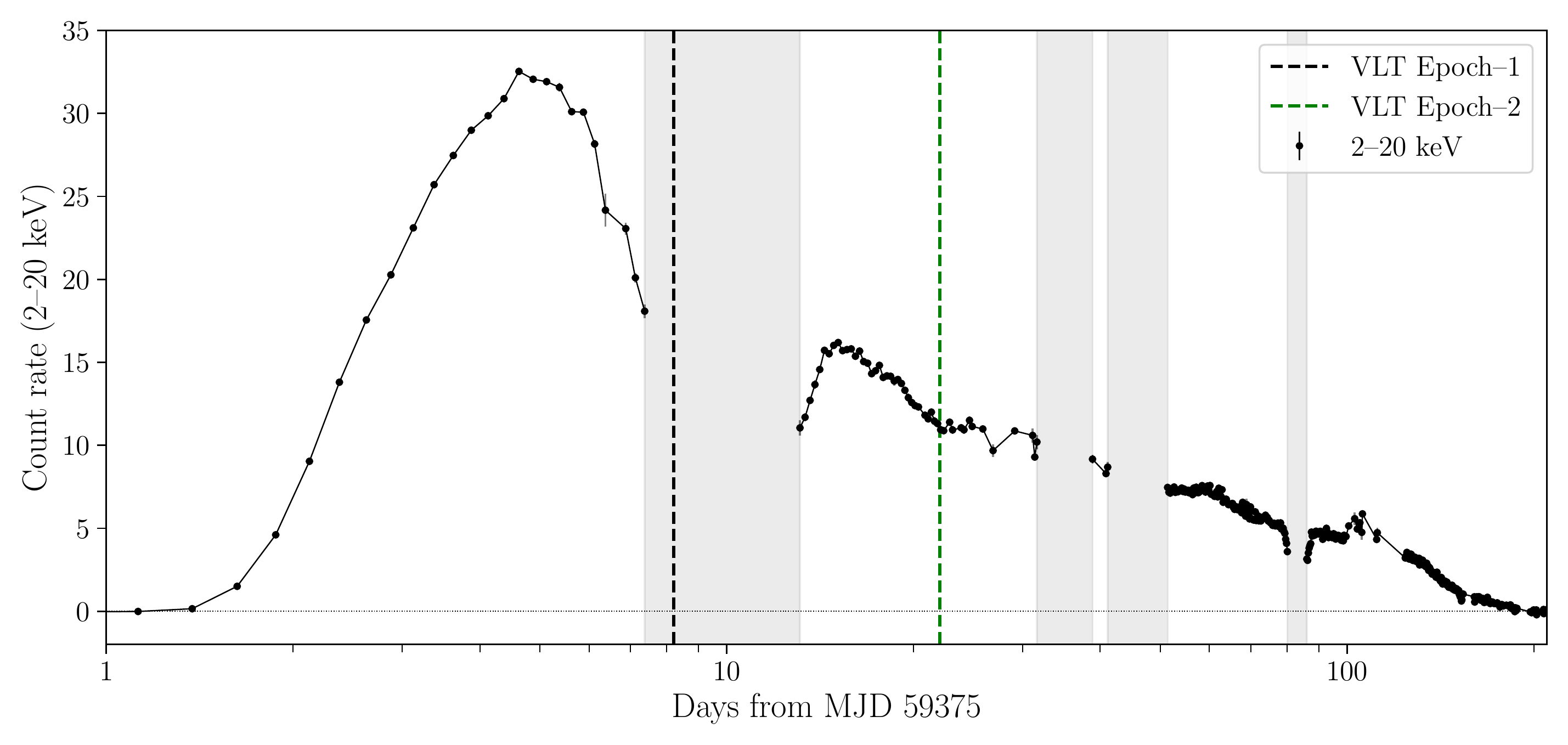}
    \caption{X-ray light curve of 4U 1543--47 during its 2021 outburst (from \textit{MAXI} data). VLT/X-shooter epochs are marked as vertical dashed lines. The grey-shadowed regions indicate \textit{MAXI} observing gaps.}
    \label{figLC}
\end{figure*}

In addition to jets, which are best seen at radio waves (e.g. \citealt{Mirabel1999, Fender2006, Gallo2018}), wind-type outflows are also commonly observed during the different stages of the outburst. Hot accretion disc winds are detected in X-rays during soft states \citep{Ueda1998,Neilsen2009,Ponti2012,Ponti2014,DiazTrigo2016}, while colder, optical winds have been observed during hard states \citep{Munoz-Darias2016,Munoz-Darias2017,Munoz-Darias2018,Munoz-Darias2019, Cuneo2020}. Furthermore, near-infrared signatures of these cold winds have been witnessed during both soft and hard states \citep{Sanchez-Sierras2020} and simultaneous optical and ultraviolet winds have been observed in the neutron star transient Swift~J1858.6--0814 (\citealt{Munoz-Darias2020, CastroSegura2022}). A possible explanation for this complex observational picture is that XRB winds are multi-phase in nature, as the properties of the wind detected in the BH transient V404 Cyg suggest (see \citealt{Munoz-Darias2022}). In this scenario, the detection of the wind throughout the outburst and across the electromagnetic spectrum changes according to the physical properties of the ejecta. However, this picture needs to be shaped by more observational data covering different types of systems (e.g. transients with short and long orbital periods) at different luminosities and orbital inclinations. 

4U 1543--47 was discovered in 1971 \citep{Matilsky1972,Li1976} and classified as a BH candidate soon after its second observed outburst in 1983 \citep{Kitamoto1984,VanderWoerd1989}. Optical spectra obtained during quiescence revealed an A2V spectral type companion star, which is an unusually early-type donor among transient BH XRBs \citep{Harmon1992}. \citet{Orosz1998} measured an orbital period of $P=1.123\pm0.008$ days and a semi-amplitude radial velocity for the donor star ($K_{2}$) of $124 \pm 4~\mathrm{km}~\mathrm{s}^{-1}$. The analysis of the photometric ellipsoidal modulation pointed to a low binary inclination, within the range $24^{\circ} \leq i \leq 36^{\circ}$. Based on this, the primary mass was constrained to $2.7 \leq M_{1} \leq 7.5~M_{\odot}$, indicating that the compact object is probably a BH. These authors estimated that the X-ray luminosity peak might have exceeded the Eddington luminosity during the 1983 outburst. They reported $3.1\lesssim L_{\mathrm{peak}}/L_{\mathrm{Edd}}\lesssim8.7$, assuming a distance of $d = 9.1 \pm 1.1~\mathrm{kpc}$, which is consistent with $d = 7.5 \pm 0.5~\mathrm{kpc}$ proposed by \citet{Jonker2004}.

On 11 Jun 2021, 4U 1543--47 entered into its first outburst since 2002 \citep{Lewis2006,Saikia2021}, following an atypically long quiescent period for this object, which had shown outbursts every $\sim$10 years since 1971. Perhaps related to this, the X-ray flux peak was twice that of 2002, approaching $L\mathrm{_{Edd}}$ \citep{Negoro2021} and reaching an ultra-luminous state (see Sect. \ref{SecResults}). The outburst lasted more than 200 days, and an optical magnitude of $V=13.90$ was reported when the X-ray emission reached its maximum \citep{Saikia2021}. This represents an outburst amplitude of $\sim 2.7~\mathrm{mag}$ relative to the quiescent magnitude, reported by \citealt{Orosz1998} ($\bar{V} \sim 16.6$).

Before 2021, four different outbursts had been observed; however, to the best of our knowledge, no optical or near-infrared spectra taken during outburst have been published. Here, we present simultaneous optical and near-infrared (OIR) spectroscopy of 4U~1543--47 obtained during its 2021 ultra-luminous state.

%--------------------------------------------------------------------
%--------------------------------------------------------------------

%-------------------------------------------------------------------
%- OBSERVATIONS AND DATA REDUCTION ---------------------------------
%-------------------------------------------------------------------

\section{Observations and data reduction}
\label{secObservationsReduction}

\begin{figure*}
    \centering
    \includegraphics[width=15truecm]{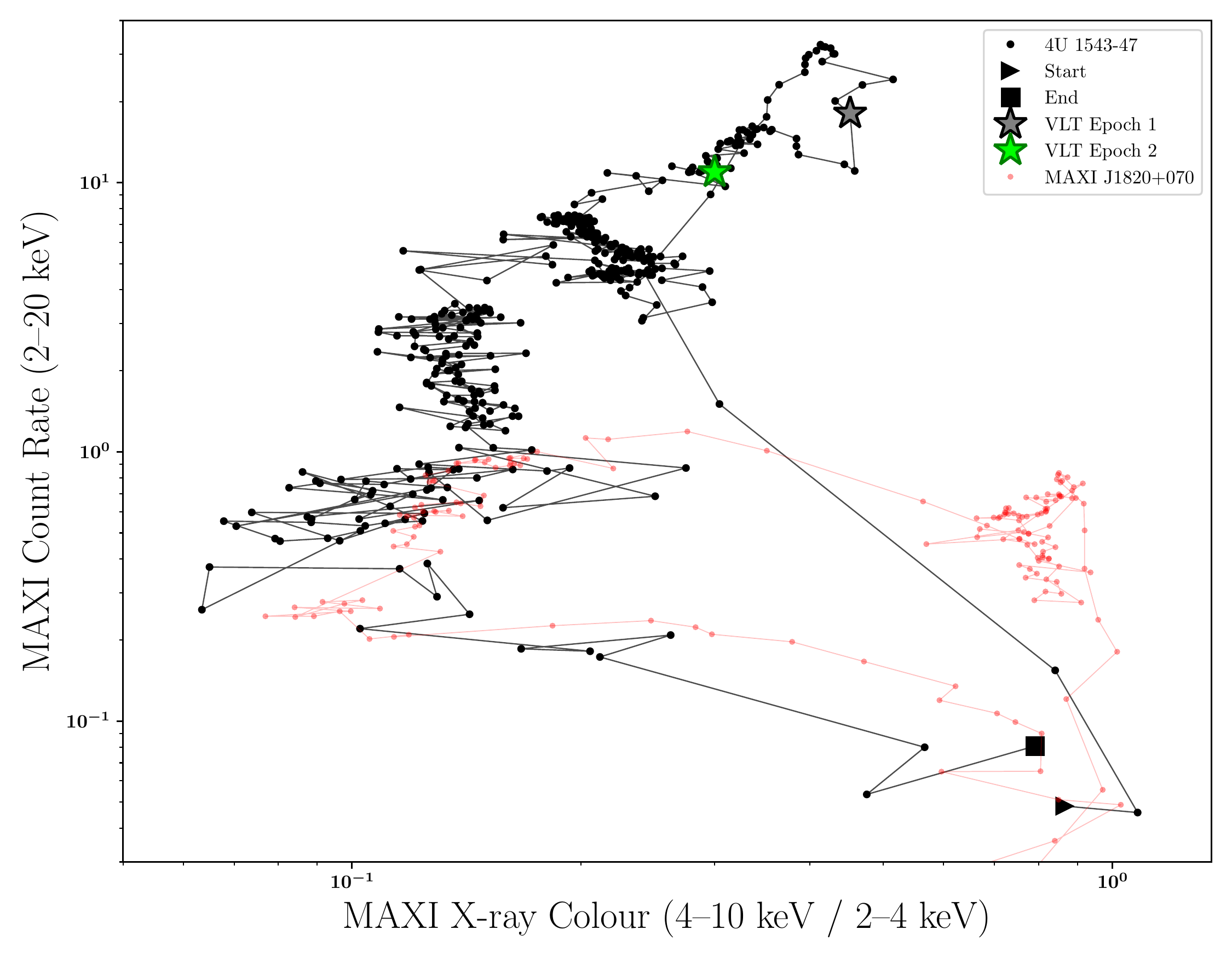}
    \caption{Hardness-intensity diagram of 4U 1543--47 during its 2021 outburst (from \textit{MAXI} data). The VLT/X-shooter spectroscopic epoch 1 and 2 are marked with grey and green stars, respectively. The evolution of the 2018-2019 outburst of MAXI J1820+070 is over-plotted in red for the sake of comparison (see text).}
    \label{figHID}%
\end{figure*}

We present two epochs of spectroscopy performed on MJD 59383.22 (18-06-2021) and MJD 59397.08 (02-07-2021) with the X-shooter spectrograph \citep{Vernet2011} attached to the Very Large Telescope (VLT; Cerro Paranal, Chile). This instrument has three arms that are operated simultaneously: UVB (3000--5560 \AA), VIS (5560--10200 \AA), and \mbox{NIR (10200--24800 \AA}). We obtained two exposures per epoch in the UVB/VIS arms with exposure times of 1162 s and 1135 s, respectively, and eight exposures of 300 s in the NIR arm. We used an ABBA nodding configuration to carry out the sky subtraction, with a total integration of 2270--2400 s per epoch and per arm. We used a slit angle of 35º and slit widths of 1.0 arcsec (UVB) and 0.9 arcsec (VIS-NIR), which rendered velocity resolutions of $\sim55~\mathrm{km}~\mathrm{s}^{-1}$ (UVB), $\sim34~\mathrm{km}~\mathrm{s}^{-1}$ (VIS), and $\sim54~\mathrm{km}~\mathrm{s}^{-1}$ (NIR). We note that the seeing was 1.43 arcsec and 0.65 arcsec during epochs -1 and -2, respectively, and both nights were classified as `clear' (i.e. non-photometric). We also used one X-shooter observation of an A2V standard star (HD88955) from the ESO data archive to evaluate the contribution of the secondary star to the total flux of the system. This was taken on MJD 58147.06 with the same instrument configuration as our science target.

The data were reduced using the X-shooter ESO Pipeline v3.5.0 as follow. We performed bias and flat-field correction. Subsequently, the centre of each order was determined and a preliminary wavelength solution was obtained (physical model) by using arc-lamp emission lines. This calibration was optimised with an alignment correction performed through the Active Flexure Compensation files, which were obtained just before each observation. This typically results in a sub-pixel accuracy (mean standard deviation of the expected line position versus real position $\lesssim 0.2$ pixels). We determined and corrected the instrument response with a standard star observed on the same night for every epoch (flux calibration procedure of the pipeline). This task takes into account the parameters of the standard star and the science object, correcting for the differences of exposure time, air mass, and pixel integration binning. Finally, we extracted the science spectra with the \textit{localization} method of the \textit{scired} X-shooter pipeline recipe, and carefully set the parameters of the cosmic-ray rejection algorithm \citep{VanDokkum2001}.

All the spectra were corrected from telluric absorption lines using \textsc{Molecfit} v3.0.3 tailored for the VIS/NIR data and v1.5.9 for the UVB data, because v3.0.3 no longer allows one to correct the latter \citep{Smette2015}. The high signal-to-noise ratio in the UVB data allowed us to fit and correct the absorption of the O$_{3}$ molecule by setting two masks of 60~\AA~in the Huggins bands (\mbox{3050--3400 \AA}), avoiding the emission lines present in this region. Since the observations with the three arms had similar exposure times, we fixed the O$_{3}$ column density value to that obtained from the UVB data when processing the VIS spectra, which were also corrected for telluric H$_{2}$O and O$_{2}$ absorption. For this, we set between five and seven masks across the spectral range covered by the VIS arm with sizes between \mbox{80 \AA}~and \mbox{120 \AA}. These masks were placed in regions where these molecules have absorption features of low to medium intensity, avoiding bad pixels and the emission features of the science spectra. Next, we fixed the values of the H$_{2}$O and O$_{2}$ column densities obtained in the VIS arm when fitting the NIR spectra. We set between four and six masks to fit the CO$_{2}$ and CH$_{4}$ absorption features. For all arms we iterated the fitting process by setting each molecular column density obtained in one cycle as the initial value for the next one.

We performed aperture photometry of 4U 1543--47 on the acquisition images of the two epochs that was calibrated against Skymapper \citep{Onken2019}. The images were taken with the X-shooter A\&G Camera just before the spectroscopic observations. We used the $r\_prime$ filter and an exposure time of 10 s. The epoch-1 and epoch-2 air masses were 1.236 and 1.087, respectively. We obtained $r_{\mathrm{ep1}} = 13.98 \pm 0.02$ and $r_{\mathrm{ep2}} = 14.27 \pm 0.02$. The data analysis was performed using the spectral analysis tool \textsc{Molly} and custom routines developed under \textsc{Python} 3.7. We built the hardness-intensity diagram and the X-ray light curve using data in bins of 6h from the Monitor of All-Sky X-ray Image (\textit{MAXI}; \citealt{Matsuoka2009})\footnote{http://maxi.riken.jp/top/index.html}.

%-------------------------------------------------------------------
%- RESULTS AND ANALYSIS --------------------------------------------
%-------------------------------------------------------------------

\begin{figure*}
    \centering
    \includegraphics[width=16truecm]{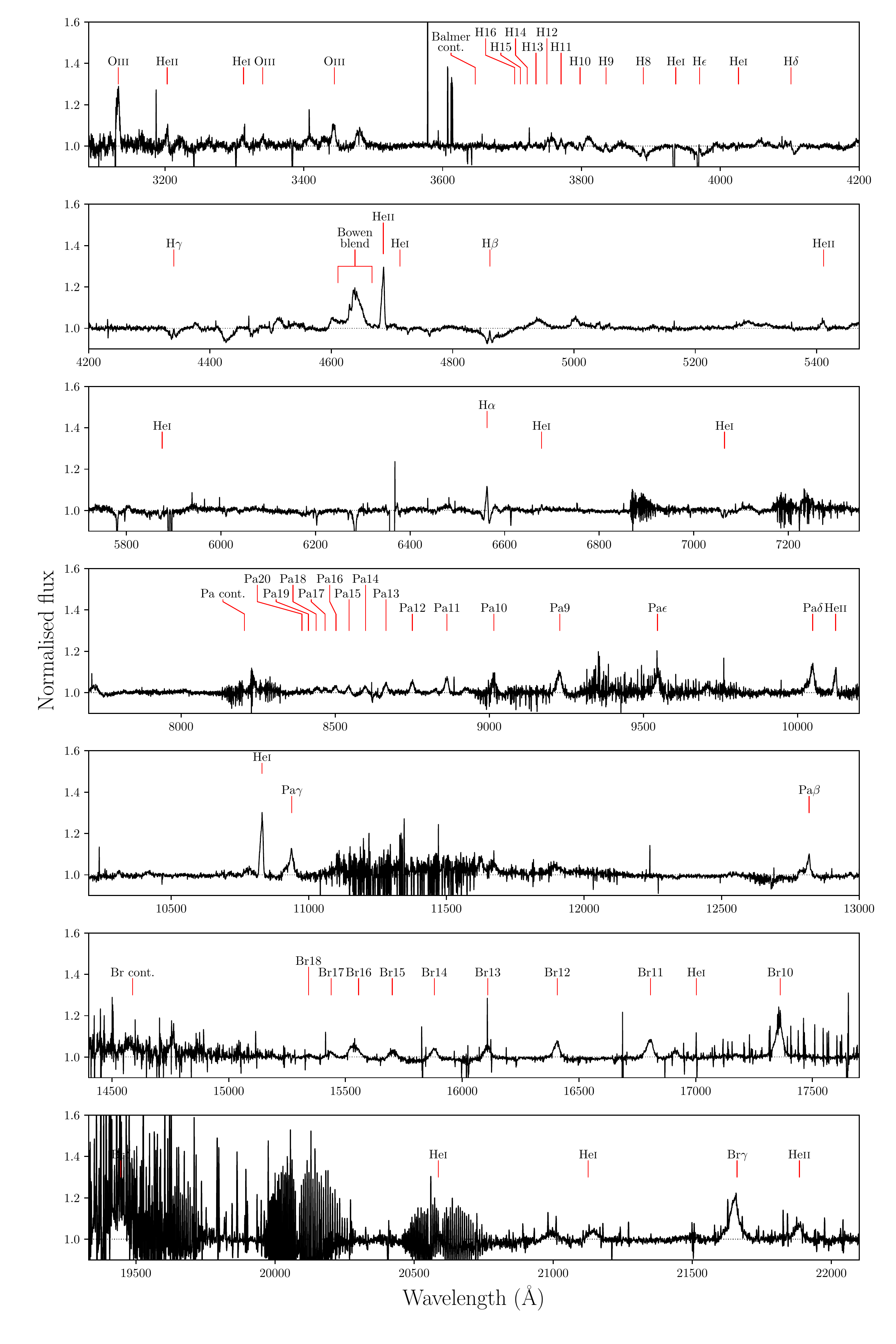}
    \caption{Normalised VLT/X-shooter spectrum of epoch 1. Major emission line features are marked.}
    \label{figEpoch1}
\end{figure*}

\section{Analysis and results}
\label{SecResults}

Figure \ref{figLC} presents the X-ray light curve of 4U~1543--47 during its 2021 outburst, while Fig. \ref{figHID} shows the hardness-intensity diagram (HID; \citealt{Homan2001}). We overplot (in red) the HID evolution of MAXI J1820+070 during its 2018-2019 outburst for a direct comparison with a standard and well-studied outburst monitored with the same instrument (e.g. \citealt{Shidatsu2019}). The MAXI J1820+070 data have been re-scaled in the vertical axis to account for the larger distance to 4U~1543--47. For this, we used $d = 2.96 \pm 0.33~\mathrm{kpc}$ \citep{Atri2020} as the distance to MAXI~J1820+070. We note that they have similar extinction along the line of sight, with relatively low column density values (\mbox{$N_{\mathrm{H}} = 4.4 \times 10^{21}~\mathrm{cm}^{-2}$} and \mbox{$N_{\mathrm{H}} = 1.5 \times 10^{21}~\mathrm{cm}^{-2}$} for 4U~1543--47 and MAXI J1820+070; see \citealt{Connors2021} and \citealt{uttley2018}).

Following a fast hard-to-soft transition (see Fig. \ref{figHID}), 4U 1543--47 moved to the soft state early in the outburst, when it dramatically increased its luminosity to become the brightest BH transient ever observed by the \textit{NICER} telescope \citep{Connors2021}. This luminous stage, which approached $L\mathrm{_{Edd}}$ for a stellar mass BH \citep{Negoro2021a}, makes a branch of its own in the HID, very similar to that seen in GRO J1655--40 during its ultra-luminous state \citep{Motta2012,Balakrishnan2020}. The two VLT/X-shooter epochs (marked by black and green stars in Figs. \ref{figLC} and \ref{figHID}) were taken during this bright phase. After the outburst peak, the system evolved to the softer region of the HID as the flux gradually decreased. Its count rate at the peak of the outburst was more than one order of magnitude higher than the scaled count rate of MAXI J1820+070, which reached $L=0.15\pm0.03~L_{\mathrm{Edd}}$ \citep{Atri2020}. Thus, 4U~1543--47 approached (and possibly exceeded) $L \sim L_{\mathrm{Edd}}$ during its 2021 outburst. This is also consistent with the soft-to-hard transition occurring at 1--4 per cent $L_{\mathrm{Edd}}$, as it is typically seen in BH transients (e.g. \citealt{Maccarone2003,Dunn2010}).

\subsection{The optical and near-infrared spectrum}
\label{secOIRSpectrum}

The two VLT/X-shooter spectra were obtained on days 8 (epoch 1) and 22 (epoch 2), where day 1 corresponds to the X-ray trigger of the outburst (MJD 59375; 10 Jun 2021). They are separated by $\sim 0.38$ in orbital phase (using the $P = 1.123 \pm 0.008~\mathrm{days}$ reported in  \citealt{Orosz1998}). The normalised spectrum of epoch 1 is shown in Fig. \ref{figEpoch1} (see Fig. \ref{figEpoch2} for the epoch-2 spectrum). In Table \ref{tableLines}, we provide the equivalent widths of the 33 emission lines unambiguously identified in both spectra. We used masks of $\pm (1500-2000)~\mathrm{km} ~\mathrm{s}^{-1}$ centred in each line, which have been slightly modified in some cases to avoid contamination by nearby spectral or instrumental features. For completeness, we also provide emission line fluxes. However, we note that although the absolute flux calibration provided by the X-shooter pipeline can be enough for some science cases, it can also be affected by systematic effects. In addition, we note that some of the emission components, especially in the Balmer lines (see below), are embedded within broader absorption features, which can modify the values of the emission components.

We found that spectra of epoch 1 and epoch 2 are quite similar. Thus, while describing the data, we mainly focus on the former and highlight the differences observed in the latter.

\subsubsection{Lower excitation emission lines}

The spectrum of 4U~1543--47 exhibits strong low excitation emission lines, such as the Balmer and Paschen hydrogen series, as well as \ion{He}{i} transitions, which are single-peaked. The Balmer emission lines from H$\alpha$ to the Balmer continuum can be observed in the three top panels of Fig. \ref{figEpoch1}. Most of these lines are embedded in broad absorption wings. Figure \ref{figDetailLines} (top-right panel) shows the H$\alpha$ profile in detail for both epochs, where a broad absorption and an emission component become apparent (see Sect. \ref{secContribSecondaryMethods}). In epoch 1, the emission lines are slightly skewed towards the red, which might indicate the presence of absorption on the blue side of the line (e.g. \hei--10830 in Fig. \ref{figDetailLines}, middle right). 

The Paschen and Brackett series are present in emission. We note that Pa$\beta$, Pa$\gamma$, and Pa$\delta$ exhibit complex profiles with intense blue emission wings (see Fig. \ref{figDetailLines}). For instance, the blue wing of the asymmetric Pa$\delta$ profile meets the continuum at $\sim$ --1200 km s$^{-1}$, while the red wing does it at $\sim$ 600 km s$^{-1}$. Contrary to the Balmer lines, the Paschen and Brackett transitions do not show any sign of underlying broad absorption features.

At longer wavelengths, we can easily identify the Brackett emission lines, from Br$\gamma$ to Br18. We note that Br$\gamma$ shows a blue wing emission excess (Fig. \ref{figDetailLines}, bottom panels), similar to that seen in the low order Paschen transitions (i.e. Pa$\beta$, Pa$\gamma$, and Pa$\delta$). In Fig. \ref{figDetailLines} (bottom-right panel), we can see Pa$\delta$ along with Br$\gamma$, which also shows a broad blue wing reaching the continuum at \mbox{$\sim$ --1200 km s$^{-1}$}. Both lines exhibit very similar blue and red emission wings, but the blue part of the Br$\gamma$ core component seems to have more intense emission.

In agreement with observations of other BH transients (see Sect. 4.2 in \citealt{Sanchez-Sierras2020}), \hei--10830 is the most intense emission line. Furthermore,  \hei--5876, \hei--6687, and \hei--7065 are also present in emission, but they are much weaker than typically observed (e.g. \citealt{MataSanchez2022}). Since these lines are embedded in a broad absorption component, their weakness could be caused by the companion star contribution (see Sect. \ref{secContribSecondaryMethods}).

\subsubsection{Higher excitation emission lines}

4U~1543--47 displays a hot OIR spectrum, with intense high excitation emission lines, such as \ion{He}{ii} and \ion{O}{iii}. This is consistent with what is observed in other high-luminosity sources, such as persistently luminous neutron stars (i.e. Z sources; see e.g. \citealt{Schachter1989} for a comprehensive analysis of the \ion{O}{iii} lines; see also \citealt{Bandyopadhyay1999} for detections of \ion{He}{ii}--21885).

\begin{figure*}[h!]
    \centering
    \includegraphics[width=17truecm]{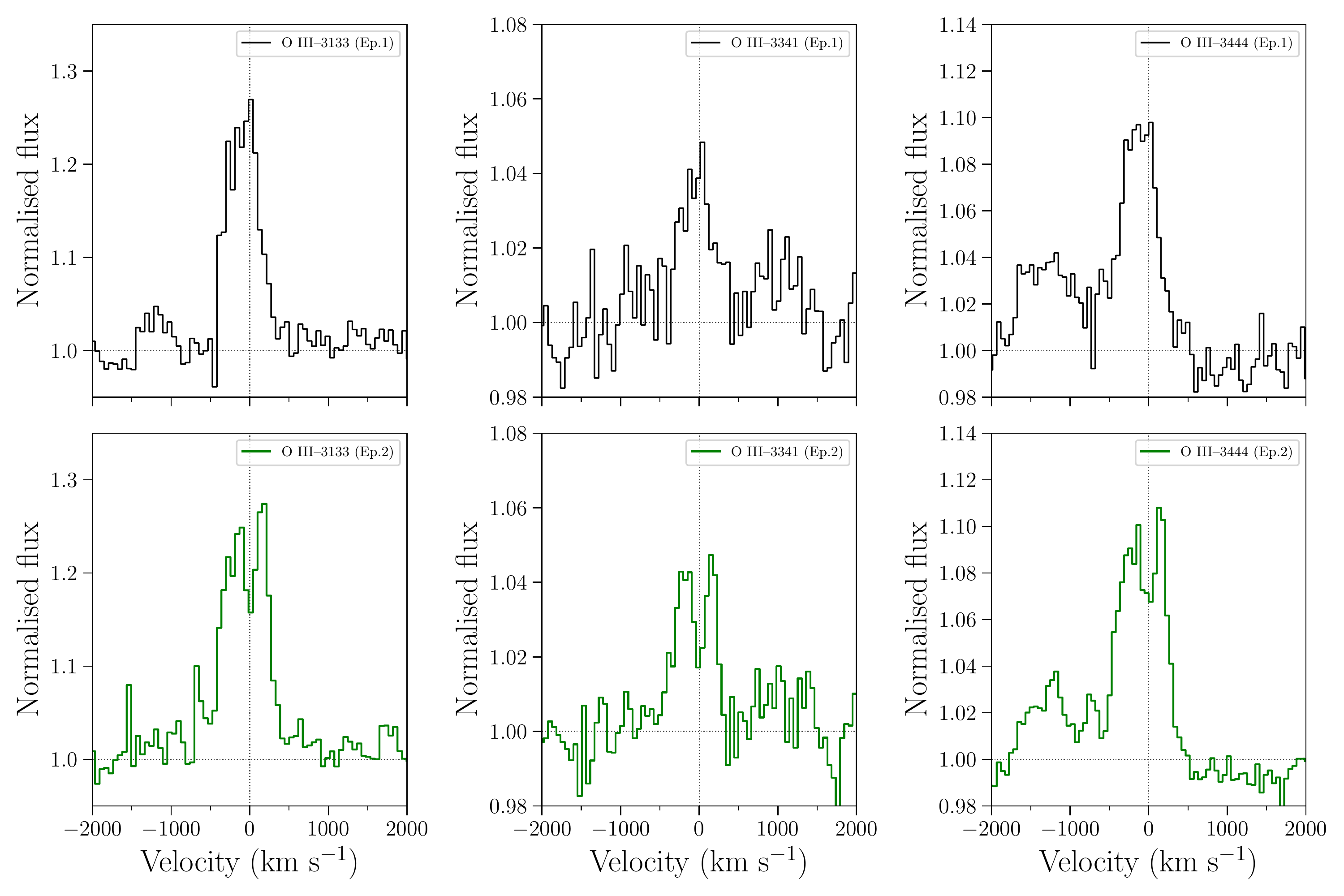}
    \caption{Details of the three \ion{O}{iii} emission lines at 3133 \AA, 3341 \AA,~and 3444 \AA~in the two epochs. The top row corresponds to epoch 1 (black) and the bottom row to epoch 2 (green). \ion{We note that O}{iii} lines are mostly single-peaked during epoch 1, whilst they are double-peaked in epoch 2.}
    \label{figDetailOIII}
\end{figure*}

In the top panel of Fig. \ref{figEpoch1}, we observe the conspicuous presence of \ion{O}{iii} emission lines at 3133~\AA, 3341~\AA,~and 3444~\AA. A close-in view of their line profiles is shown in Fig. \ref{figDetailOIII}. The lines are double-peaked during epoch 2 but not during epoch 1 (i.e. at higher luminosity). The separation between the two peaks is $353 \pm 13~\mathrm{km~s^{-1}}$, which was calculated by performing a two-Gaussian fit to the \ion{O}{iii} lines in epoch 2. This value is consistent with those typically found in other low-inclination systems (see \citealt{Panizo-Espinar2022} for a discussion on the topic). The presence of a double peak in \ion{O}{iii} represents a significant difference with respect to the H and He lines, which are single-peaked in both epochs, including the intense \heii--4686 and \heii--10124 lines. Other \heii~lines present in the spectra are \heii--3203, \heii--5411, and \heii--21885.

\begin{figure*}[h!]
    \centering
    \includegraphics[width=16truecm]{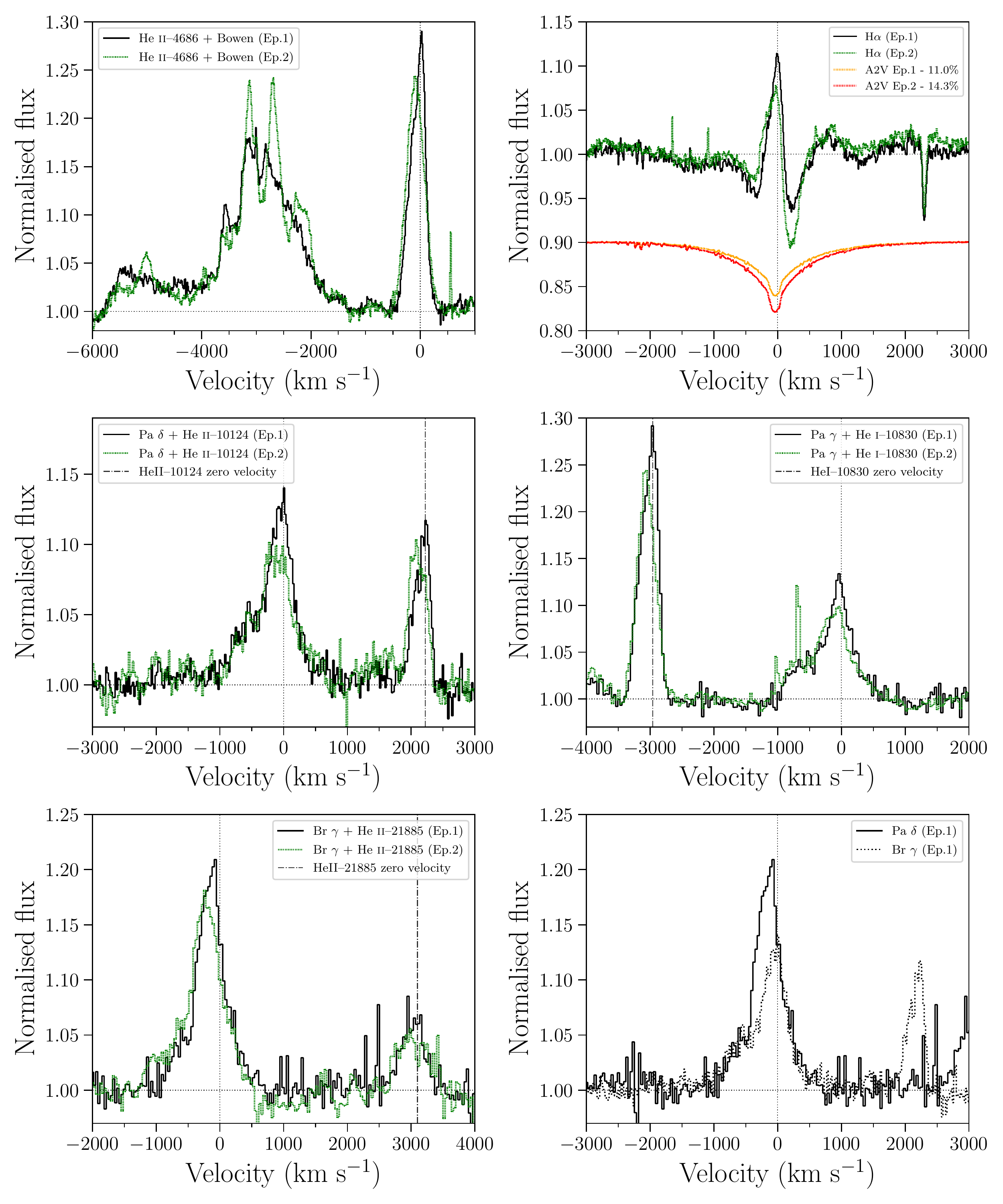}
    \caption{Details of some emission lines in the spectra of both epochs. Zero velocities of \heii--10124, \hei--10830, and \heii--21885 are marked with dash-dotted lines. We note that all H and He lines display clear red skewed profiles during epoch 1. \textit{Top left}: \heii--4686 + Bowen blend. The Bowen blend exhibits narrow emission components during epoch 2, which are not present in epoch 1. \textit{Top right}: H$\alpha$ profile. The emission is embedded in a broad absorption feature. The normalised  spectra of an A2V star, with a scale-corrected contribution, are shown vertically displaced at 0.9 for comparison. \textit{Middle left}: Pa$\delta$ emission line and \heii--10124 at $\sim$ 2200 km s$^{-1}$. We note that Pa$\delta$ shows a strong emission blue wing compared with the\heii~line. \textit{Middle right}: Pa$\gamma$ and \hei--10830 lines. We note that Pa$\gamma$ exhibits a blue emission wing. \textit{Bottom left}: Br$\gamma$ and \heii--21885 lines. \textit{Bottom right}: Pa$\delta$ (solid) and Br$\gamma$ (dotted) emission lines in epoch 1.}
    \label{figDetailLines}
\end{figure*}

As expected from the presence of the above high excitation transitions, the Bowen blend \citep{McClintock1975} is also intense (Fig. \ref{figDetailLines}, top-left panel). These fluorescence lines show three narrow emission peaks of C~\textsc{iii} (4647~\AA~\& 4650~\AA, appearing blended) and N~\textsc{iii}~(4634~\AA~\& 4641~\AA). They are particularly narrow and intense in epoch 2 (green dotted line in Fig. \ref{figDetailLines}). We note that these sharp components are consistent with reprocessed emission from the irradiated side of the donor, as seen in other systems (e.g. \citealt{Steeghs2002,Casares2003,Munoz-Darias2009b,Wang2017,Jimenez-Ibarra2018}).

\subsection{Contribution from the companion star}
\label{secContribSecondaryMethods}

Broad absorptions superimposed onto the Balmer emission cores are typically observed in XRBs during outburst (e.g. \citealt{Zurita2002,Jimenez-Ibarra2019a}). They might arise in a chromosphere-like optically thin region above the disc \citep{Dubus2001} or result from absorption by cold gas near the rim of the disc \citep{Soria2000}. Alternatively, in the case of luminous early-type donor stars, they can also reveal a significant contribution of this companion star to the optical light. The donor of 4U 1543--47 is an A2V star \citep{Orosz1998}, which totally dominates the optical spectrum during quiescence \citep{Chevalier1989} and could have a non-negligible contribution during outburst as well \citep{Russell2020a}. To estimate this contribution, we used the magnitudes obtained from the acquisition images of both epochs ($r_{\mathrm{ep1}} = 13.98 \pm 0.02$; $r_{\mathrm{ep2}} = 14.27 \pm 0.02$), together with the quiescence ($V_{\mathrm{qu}} = 16.56 \pm 0.01$) and the outburst ($V_{\mathrm{pk}} = 13.90 \pm 0.01$) values reported by \citealt{Saikia2021}. We converted $V$ to $r$ magnitudes through the empirical formula given by \citealt{Jordi2006} (Table 3, case $V - R \leq 0.93$), which yields $r_{\mathrm{pk}} = 13.72 \pm 0.02$. This translates into a contribution of the A2V star to the total r-band light of $8.63 \pm 0.16 \%$ at the peak of the outburst, and $11.0 \pm 0.3 \%$ and $14.3 \pm 0.4 \%$  for epochs 1 and 2, respectively. In Fig. \ref{figDetailLines} (top-right panel), we observe that the absorptions in both epochs are in the range of $\sim$ 5--10\% below the continuum level, while Balmer absorptions reach $\sim 50 \%$ in a standard A2V star. The above strongly suggests that Balmer absorption components are the result of a significant contribution of the donor to the optical spectrum. Given the early spectral type, this contribution sharply drops towards the near-infrared, which explains the lack of similar broad absorption features in such a spectral range.

%-------------------------------------------------------------------
%- DISCUSSION ------------------------------------------------------
%-------------------------------------------------------------------
\section{Discussion}
\label{secDiscussion}

We have presented two X-shooter spectra taken during the 2021 outburst of the BH transient 4U 1543--47, which we put in the context of the outburst evolution of the source using X-ray data from \textit{MAXI}. These are, to the best of our knowledge, the first OIR spectra of a BH transient obtained during an ultra-luminous state.

We found that the spectra are very rich in the presence of emission lines, especially in low excitation lines, which is consistent with other BH transients in outburst (e.g. \citealt{Munoz-Darias2019,Cuneo2020}). However, some higher excitation emission lines are particularly interesting in 4U 1543--47 as compared with other sources. For instance, intense \ion{O}{iii} lines with variant profiles are present (see Sect. \ref{secOIIIdiscussion} for further discussion), as opposed to other systems where they are very weak or absent (see e.g. MAXI J1803--298, \citealt{MataSanchez2022} and MAXI J1348--630, \citealt{Panizo-Espinar2022}). Likewise, \ion{He}{ii}--21885, which is intense in both epochs, is absent in other BH transients during outburst (see also GX 339--4, \citealt{Rahoui2014}, or MAXI J1820+070, \citealt{Sanchez-Sierras2020}). This could be related to the high luminosity of 4U 1543--47, which likely reached the Eddington luminosity, while most BH transients only rise up to a few per cent of $L_{\mathrm{Edd}}$. In addition, this would also explain the similarities (presence of high excitation emission lines) with the spectra of other high-luminosity systems, such as the Z sources \citep{Bandyopadhyay1999,Steeghs2002,MataSanchez2015}.

The complex line profiles of some near-infrared transitions are comparable to those seen in other BH transients during outburst (e.g. \citealt{Sanchez-Sierras2020,Panizo-Espinar2022}). These profiles include features usually related to the presence of an outflow, which is discussed in the following section.

\subsection{Tentative wind-related features}

We do not observe definite wind signatures, such as P-Cygni profiles \citep{Castor1979}, in our spectra. However, some of the features found in the OIR line profiles of 4U 1543--47 could be interpreted as signatures of outflows:

First, the red skewed profiles observed in the low-excitation lines during epoch 1 (see Fig. \ref{figDetailLines}) could be explained by the presence of a low-velocity outflow, as previously suggested for other sources (\citealt{Munoz-Darias2019,Panizo-Espinar2022}). Since the X-ray flux during epoch 1 is twice that of epoch 2 (see Figs. \ref{figLC} and \ref{figHID}), the presence of these putative outflow features could be related to the stronger radiation field. However, we note that although the outflow has been suggested to be permanently active in BH XRBs during outburst, its observability (and thus detectability) depends on the physical conditions of the ejecta. These are sensitive to several factors, such as the local radiation field (see \citealt{Munoz-Darias2022} and references therein for a discussion on this topic). An alternative explanation for the asymmetric profiles might be an anchored structure in the accretion flow, such as a hot spot (i.e. the bulge). However, we note that the $\sim 0.38$ difference in orbital phase between the two epochs goes against this scenario.

Secondly, blue, broad emission line wings are observed in both epochs, and they are commonly associated with outflows, not only in XRBs \citep{Rahoui2014,Sanchez-Sierras2020,Cuneo2020,Panizo-Espinar2021} but also in supernovae \citep{Borish2015}, young stellar objects \citep{Antoniucci2017} and Seyfert galaxies \citep{Bing2019}. Although these features are consistent with the presence of outflows, we note that they do not constitute a definitive piece of evidence. Related to this, \citealt{Fender1999} proposed that a flattened disc wind could explain similar profiles found in the near-infrared emission lines of another very luminous source, Cyg X-3.

Features that constitute strong evidence for the presence of an outflow, such as P-Cygni profiles, are commonly found in BH transients with high inclinations (e.g. \citealt{Jimenez-Ibarra2019a}). Since 4U 1543--47 is a low-inclination system, the detection of an outflow could be affected. For instance, due to the additional presence of emission lines arising in the accretion disc, the presence of a low-velocity absorption could simply result in asymmetric line profiles (skewed towards the red) instead of conspicuous blue-shifted absorption components lying below the continuum level. The above are detected in some Paschen and Brackett lines (see Fig. \ref{figDetailLines}) and have also been observed in the low-to-mid inclination BH transient MAXI J1348--630, and have been associated with outflow signatures \citep{Panizo-Espinar2022}.

The detection of outflows also seems to depend on the accretion state. Cold winds have mainly been detected  in the optical lines of BH transients during the hard state (e.g. \citealt{Munoz-Darias2019}). However, these winds have also been observed in near-infrared lines during the soft state (e.g. \citealt{Sanchez-Sierras2020}). If the presence of winds in 4U 1543--47 were confirmed, this would represent an additional step towards strengthening the global picture of XRB winds that we are building (see \citealt{Panizo-Espinar2022} for fully summarised wind detections on transients). We note that our OIR spectra of 4U 1543--47 are the first obtained ones during the poorly studied ultra-luminous state. The tantalising evidence of outflow signatures, along with the wind detection in other low-inclination systems, would strengthen the idea of a persistent wind during the outburst of XRB transients \citep{Munoz-Darias2016,Sanchez-Sierras2020,CastroSegura2022,Tetarenko2018,Dubus2019,Petrucci2021}. Nevertheless, more observations during these uncommon states are needed to achieve a firm conclusion.

As already noted, the branch of the HID (see Fig. \ref{figHID}) in which our spectra were taken is reminiscent of that shown by the BH transient GRO J1655-40 during its ultra-luminous state. This system is also well known for showing conspicuous X-ray winds, for which different launching mechanisms have been proposed (see e.g. \citealt{Miller2006, Shidatsu2016, Neilsen2016}). These (\textit{Chandra}) wind detections occurred on MJD 53461, several days before the system entered the ultra-luminous branch (MJD 53500; \citealt{Motta2012}), and thus a direct comparison with our results is not straightforward.

\subsection{The \ion{O}{iii} emission lines}
\label{secOIIIdiscussion}

The high-luminosity outburst of 4U 1543--47 offered an excellent opportunity to gather high-quality multi-wavelength spectra and, thereby, to study the elusive \ion{O}{iii} emission lines, whose location in the bluest part of the optical spectrum usually prevents their observation. These \ion{O}{iii} transitions play an important role in the Bowen fluorescence emission process, which is a result of a blend of resonance lines of \ion{He}{ii}, \ion{N}{iii}, and \ion{O}{iii} \citep{McClintock1975}. Interestingly, the \ion{O}{iii} line profiles significantly change between epochs, from single-peaked (epoch 1) to double-peaked (epoch 2; see Fig. \ref{figDetailOIII}). Double-peaked profiles are associated with moving material in an accretion disc \citep{Horne1986}. Since 4U 1543--47 is a low inclination system \citep{Orosz1998}, the epoch-2 double-peaked lines could arise from an inner region of the accretion disc, where the projected radial velocity is large enough to allow us to resolve the two peaks. Thus, in epoch 1, at higher luminosity, the emission would arise in an outer disc (lower-velocity) region, where a double peak cannot be discerned. Alternatively, the single-peaked lines observed in epoch 1 could partially originate in an outflow, seen as an additional emission component that would fill in the double peak. This would be in agreement with other possible outflow-related features found during this epoch (see above) and with theoretical studies suggesting an outflow origin for the single-peaked emission lines found in accreting white dwarfs \citep{Matthews2015}.

%-------------------------------------------------------------------
%- CONCLUSIONS -----------------------------------------------------
%-------------------------------------------------------------------
\section{Conclusions}
\label{secConclusions}

We have presented for the first time simultaneous OIR spectroscopy of 4U 1543--47 during outburst, which provides unique insight into the OIR spectrum of BH transients during the ultra-luminous state. The spectra show a hot accretion disc, with intense high excitation emission lines, such as \ion{He}{ii} or \ion{O}{iii}. These are similar to those found in other high-luminosity sources, such as Z sources (i.e. persistently luminous neutron stars). In addition, the broad absorption features in Balmer transitions that we observe are most likely produced by the contribution of the A2V donor star.

We do not find definitive signatures that imply the presence of an outflow in this system, such as those observed in other BH transients. However, other observables, such as red skewed emission profiles and blue broad emission wings, could be interpreted as signatures of an outflow in the system. Its low inclination could affect the visibility of some of these signatures. Future observations of 4U 1543--47 or another BH transient displaying an ultra-luminous state may shed some light on the properties and relevance of outflows during this accretion state.

\begin{acknowledgements}
We are thankful to the anonymous referee for constructive comments that have improved this paper. We acknowledge support from the Spanish Ministry of Science and Innovation via the \textit{Europa Excelencia} program EUR2021-122010 and the \textit{Proyectos de Generación de Conocimiento}  PID2020-120323GB-I00 and PID2021-124879NB-I00. We acknowledge support from the \textit{Consejería de Economía, Conocimiento y Empleo del Gobierno de Canarias} and the European Regional Development Fund under grant with reference ProID2021010132 ACCISI/FEDER, UE. This project acknowledges funding from the European Research Council (ERC) under the European Union’s Horizon 2020 research and innovation programme (grant agreement No 865637). Based on observations collected at the European Southern Observatory under ESO programmes 105.20LK.002 and 60.A-9022(C). Federico Vicentelli acknowledges support from the grant FJC2020-043334-I financed by MCIN/AEI/10.13039/501100011033 and NextGenerationEU/PRTR. This research has made use of MAXI data provided by RIKEN, JAXA and the MAXI team. The national facility capability for SkyMapper has been funded through ARC LIEF grant LE130100104 from the Australian Research Council, awarded to the University of Sydney, the Australian National University, Swinburne University of Technology, the University of Queensland, the University of Western Australia, the University of Melbourne, Curtin University of Technology, Monash University and the Australian Astronomical Observatory. SkyMapper is owned and operated by The Australian National University's Research School of Astronomy and Astrophysics. The survey data were processed and provided by the SkyMapper Team at ANU. The SkyMapper node of the All-Sky Virtual Observatory (ASVO) is hosted at the National Computational Infrastructure (NCI). Development and support of the SkyMapper node of the ASVO has been funded in part by Astronomy Australia Limited (AAL) and the Australian Government through the Commonwealth's Education Investment Fund (EIF) and National Collaborative Research Infrastructure Strategy (NCRIS), particularly the National eResearch Collaboration Tools and Resources (NeCTAR) and the Australian National Data Service Projects (ANDS). \textsc{Molly} software developed by Tom Marsh is gratefully acknowledged.

\end{acknowledgements}

% WARNING
%-------------------------------------------------------------------
% Please note that we have included the references to the file aa.dem in
% order to compile it, but we ask you to:
%
% - use BibTeX with the regular commands:
%   \bibliographystyle{aa} % style aa.bst
%   \bibliography{Yourfile} % your references Yourfile.bib
%
% - join the .bib files when you upload your source files
%-------------------------------------------------------------------
\bibliographystyle{aa}
\bibliography{Libreria}

\begin{appendix}
\section{Fluxes and equivalent widths}

\begin{table*}[ht!]
\centering
\caption{Equivalent widths and fluxes of the most prominent lines in the two epochs.}
  \begin{threeparttable}
        \begin{tabular}{|c|c|c|c|c|c|}
        \hline
        \multirow{3}{*}{\textbf{Line}} & \multirow{3}{*}{$\lambda_{o}$ $(\mbox{\AA})$ \tnote{a}} & \multicolumn{2}{|c|}{Equivalent Width} & \multicolumn{2}{|c|}{Flux density}\\
         & & \multicolumn{2}{|c|}{(\AA)} & \multicolumn{2}{|c|}{($\times 10^{-17}$ erg cm$^{-2}$ s$^{-1}$ \AA$^{-1}$) \tnote{c} } \\
        \cline{3-6}
         & & Epoch 1 & Epoch 2 & Epoch 1 & Epoch 2 \\
        \hline
        \hline
\ion{O}{iii}--3133 & 3132.86 & 0.59 $\pm$ 0.06 & 1.34 $\pm$ 0.04 & 104 $\pm$ 11 & 352 $\pm$ 11 \\
\ion{He}{ii}--3203 & 3203.14 & 0.44 $\pm$ 0.03 & 0.37 $\pm$ 0.02 & 105 $\pm$ 9 & 135 $\pm$ 8 \\
\ion{He}{i}--3313 & 3313.25 & 0.26 $\pm$ 0.03 & 0.25 $\pm$ 0.02 & 32 $\pm$ 4 & 49 $\pm$ 4 \\
\ion{O}{iii}--3341 & 3340.74 & 0.14 $\pm$ 0.02 & 0.10 $\pm$ 0.02 & 19 $\pm$ 3 & 22 $\pm$ 4 \\
\ion{O}{iii}--3444 & 3444.10 & 0.70 $\pm$ 0.02 & 0.83 $\pm$ 0.01 & 112 $\pm$ 3 & 212 $\pm$ 4 \\
\tnote{b}~~H $\delta$ & 4101.73 & -0.20 $\pm$ 0.01 & -0.31 $\pm$ 0.01 & -11.8 $\pm$ 0.8 & -28.4 $\pm$ 1.0 \\
\tnote{b}~~H $\gamma$ & 4340.47 & -0.79 $\pm$ 0.01 & -0.59 $\pm$ 0.01 & -40.1 $\pm$ 0.6 & -46.7 $\pm$ 0.8 \\
\ion{He}{ii}--4686 & 4685.75 & 1.38 $\pm$ 0.01 & 1.53 $\pm$ 0.01 & 120.7 $\pm$ 0.6 & 208.5 $\pm$ 0.9 \\
\tnote{b}~~H $\beta$ & 4861.33 & -0.46 $\pm$ 0.01 & -0.42 $\pm$ 0.01 & -26.1 $\pm$ 0.4 & -36.4 $\pm$ 0.6 \\
\ion{He}{ii}--5411 & 5411.55 & 0.27 $\pm$ 0.01 & 0.21 $\pm$ 0.01 & 18.1 $\pm$ 0.4 & 21.6 $\pm$ 0.8 \\
\tnote{b}~~H $\alpha$ & 6562.76 & -0.09 $\pm$ 0.01 & -0.42 $\pm$ 0.01 & -1.30 $\pm$ 0.17 & -8.8 $\pm$ 0.3 \\
Pa 14 & 8598.39 & 0.65 $\pm$ 0.01 & 0.79 $\pm$ 0.01 & 4.60 $\pm$ 0.06 & 8.3 $\pm$ 0.2 \\
Pa 13 & 8665.02 & 0.98 $\pm$ 0.01 & 0.91 $\pm$ 0.01 & 6.81 $\pm$ 0.06 & 9.3 $\pm$ 0.2 \\
Pa 12 & 8750.47 & 0.66 $\pm$ 0.01 & 0.71 $\pm$ 0.01 & 3.87 $\pm$ 0.06 & 6.1 $\pm$ 0.1 \\
Pa 11 & 8862.78 & 1.07 $\pm$ 0.01 & 0.77 $\pm$ 0.01 & 8.17 $\pm$ 0.06 & 8.6 $\pm$ 0.1 \\
Pa 10 & 9014.91 & 0.77 $\pm$ 0.01 & 1.04 $\pm$ 0.02 & 4.11 $\pm$ 0.08 & 8.0 $\pm$ 0.1 \\
Pa 9 & 9229.01 & 0.91 $\pm$ 0.02 & 1.86 $\pm$ 0.01 & 3.48 $\pm$ 0.07 & 10.2 $\pm$ 0.1 \\
Pa $\epsilon$ & 9545.97 & 2.47 $\pm$ 0.02 & 2.40 $\pm$ 0.02 & 18.09 $\pm$ 0.14 & 25.15 $\pm$ 0.17 \\
Pa $\delta$ & 10049.37 & 2.50 $\pm$ 0.02 & 1.57 $\pm$ 0.02 & 8.74 $\pm$ 0.06 & 8.06 $\pm$ 0.08 \\
\ion{He}{ii}--10124 & 10123.77 & 1.01 $\pm$ 0.01 & 1.04 $\pm$ 0.01 & 10.67 $\pm$ 0.12 & 15.96 $\pm$ 0.18 \\
\ion{He}{i}--10830 & 10830.17 & 3.14 $\pm$ 0.04 & 2.81 $\pm$ 0.03 & 11.33 $\pm$ 0.14 & 14.40 $\pm$ 0.18 \\
Pa $\gamma$ & 10938.09 & 2.74 $\pm$ 0.07 & 2.92 $\pm$ 0.05 & 4.62 $\pm$ 0.12 & 7.02 $\pm$ 0.11 \\
Pa $\beta$ & 12818.08 & 2.33 $\pm$ 0.05 & 1.84 $\pm$ 0.05 & 2.48 $\pm$ 0.06 & 2.69 $\pm$ 0.08 \\
Br 16 & 15556.47 & 2.80 $\pm$ 0.03 & 3.07 $\pm$ 0.04 & 1.74 $\pm$ 0.03 & 2.53 $\pm$ 0.03 \\
Br 15 & 15700.68 & 1.78 $\pm$ 0.05 & 2.10 $\pm$ 0.06 & 0.80 $\pm$ 0.03 & 1.24 $\pm$ 0.03 \\
Br 14 & 15880.56 & 2.21 $\pm$ 0.05 & 2.17 $\pm$ 0.05 & 0.89 $\pm$ 0.03 & 1.16 $\pm$ 0.03 \\
Br 13 & 16109.33 & 1.80 $\pm$ 0.05 & 2.26 $\pm$ 0.04 & 0.72 $\pm$ 0.03 & 1.21 $\pm$ 0.03 \\
Br 12 & 16407.21 & 2.19 $\pm$ 0.05 & 3.38 $\pm$ 0.04 & 0.76 $\pm$ 0.02 & 1.55 $\pm$ 0.02 \\
Br 11 & 16806.54 & 1.03 $\pm$ 0.05 & 3.50 $\pm$ 0.05 & 0.35 $\pm$ 0.02 & 1.68 $\pm$ 0.03 \\
Br 10 & 17362.13 & 8.04 $\pm$ 0.05 & 7.73 $\pm$ 0.05 & 2.29 $\pm$ 0.02 & 2.84 $\pm$ 0.02 \\
\ion{He}{i}--21126 & 21125.83 & 2.56 $\pm$ 0.04 & 0.09 $\pm$ 0.05 & 0.48 $\pm$ 0.01 & 0.020 $\pm$ 0.012 \\
Br $\gamma$ & 21661.18 & 8.87 $\pm$ 0.07 & 9.20 $\pm$ 0.07 & 1.10 $\pm$ 0.01 & 1.44 $\pm$ 0.02 \\
\ion{He}{ii}--21885 & 21885.00 & 2.49 $\pm$ 0.06 & 2.49 $\pm$ 0.05 & 0.63 $\pm$ 0.02 & 0.80 $\pm$ 0.02 \\
        \hline
        \hline
    \end{tabular}
    \begin{tablenotes}
    \item[a]{Laboratory wavelength.}
    \item[b]{Lines strongly affected by the companion star spectrum.}
    \item[c]{It is important to note that the quoted errors are purely formal and are more likely underestimated because of slit losses due to atmospheric refraction and seeing conditions.}
    \end{tablenotes}
  \end{threeparttable}
  \label{tableLines}
\end{table*}

\section{Epoch-2 full spectrum}
\label{appendFigEpoch2}

Here, we show the fully normalised spectrum of epoch 2.

\begin{figure*}
    \centering
    \includegraphics[width=16truecm]{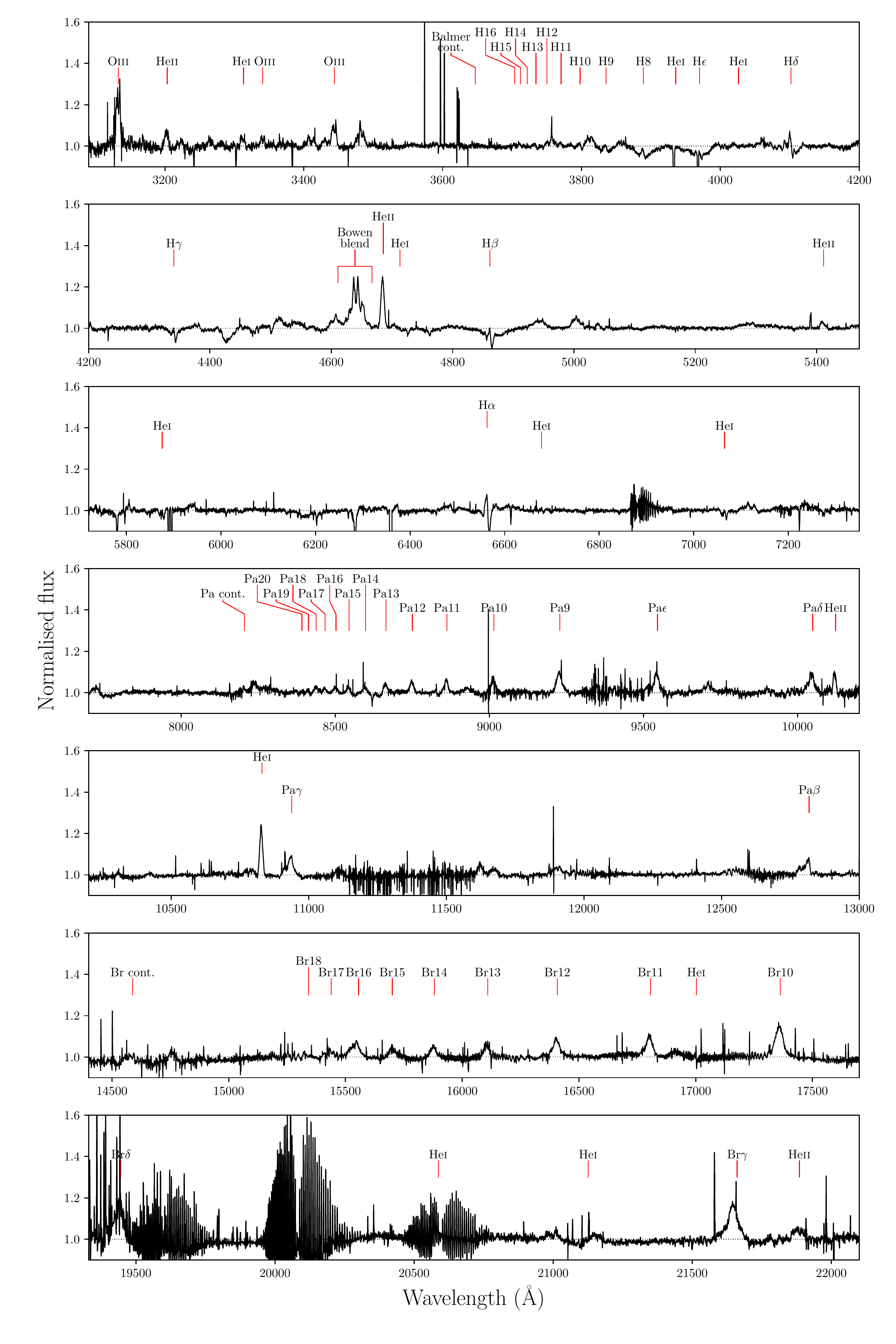}
    \caption{Normalised VLT/X-shooter spectrum of epoch 2. Major emission line features are marked.}
    \label{figEpoch2}
\end{figure*}

\end{appendix}

\end{document}